# Hypergraphic representation for adaptive quantum circuits

Waldemir Cambiucci
*Dept. of Computing Engineering and Digital Systems*
*University of São Paulo*
São Paulo, Brasil
waldemir.cambiucci@usp.br

Regina Melo Silveira
*Dept. of Computing Engineering and Digital Systems*
*University of São Paulo*
São Paulo, Brasil
regina.silveira@usp.br

Wilson Vicente Ruggiero
*Dept. of Computing Engineering and Digital Systems*
*University of São Paulo*
São Paulo, Brasil
wruggiero@usp.br

*Abstract*—Adaptive quantum circuits enhance flexibility and efficiency over traditional static circuits by dynamically adjusting their structure and parameters in real-time based on intermediate measurement outcomes. This paper introduces a novel hypergraph representation for adaptive quantum circuits, where groups of gates are considered as participants of hyperedges. By incorporating these gate groups into hyperedges, we create an extended hypergraph that includes constraints usable during the partitioning process. This approach guides the partitioning to maintain groups of ports associated with classical operations, ensuring that the resulting partitions prioritize qubits involved in the same sections of classical operations inherent to the adaptive approach. We present a new hypergraph partitioning algorithm based as an extension of Fiduccia-Mattheyses heuristic, to support hypergraphs created from adaptive quantum circuits. Comparative analysis between static and adaptive methods demonstrates the effectiveness of the proposed hypergraph techniques for adaptive circuits. Experimental results using benchmark quantum circuits validate our theoretical insights, showing improvements in circuit representation for partitioning heuristics. These findings highlight the practical benefits of hypergraph representation in adaptive quantum computing.

## I. INTRODUCTION

Quantum computing has shown remarkable promise in solving intricate computational problems [1]. However, its practical implementation is limited by the scarcity of available qubits for processing tasks, during this Noisy Intermediate-Scale Quantum (NISQ) devices [2]. Advancements in quantum hardware have introduced capabilities such as mid-circuit measurements and qubit resets, which allow qubits that have been measured to be reused [3]. This innovation significantly reduces the number of qubits necessary to efficiently execute quantum algorithms, as presented in [4]. This new field has been called as Adaptive Quantum Circuits (AQC). We also find related works using names as Dynamic Circuits [5] and Hybrid Circuit [6], considering the adaptability or dynamic behavior from these circuits. For this paper, we will be using the expression "adaptive [quantum] circuits" or AQC.

Adaptive circuits represent a significant advancement in the field of quantum computing, offering enhanced flexibility and efficiency compared to traditional, static quantum circuits. Unlike conventional static quantum circuits, which follow a predetermined sequence of quantum gates and operations, AQC can dynamically adjust their structure and parameters in response to intermediate measurement outcomes or environmental conditions. This adaptability allows for more sophisticated error correction, optimization of quantum algorithms, and improved resilience against decoherence and other quantum noise.

One of the key advantages of adaptive quantum circuits is their ability to implement feedback mechanisms within quantum computations [7]. By incorporating measurements and conditional operations, these circuits can effectively respond to the probabilistic nature of quantum states, enabling more precise control over the quantum system. This capability is particularly valuable in tasks such as quantum error correction, where real-time adjustments are essential to maintain the integrity of quantum information. Additionally, adaptive circuits can enhance the performance of variational quantum algorithms by iteratively refining parameters based on measurement results, leading to faster convergence and better optimization outcomes.

### A. Background and motivation

The development of adaptive quantum circuits also opens new avenues for exploring complex quantum phenomena and designing more robust quantum architectures. Several related works are actively investigating various strategies to integrate adaptability into quantum hardware and software, addressing challenges related to coherence times, gate fidelities, and scalability [8][9]. However, the development and implementation of adaptive quantum circuits are not without significant challenges. Optimizing these circuits requires sophisticated algorithms that can efficiently manage the dynamic adjustments in real-time, ensuring that the system remains both stable and performant under varying conditions. The complexity of such optimization tasks increases with the size of the quantum system, making scalability a critical concern. Additionally, circuit cutting, dividing a large quantum circuit into smaller, more manageable sections that can be processed independently, poses another layer of difficulty. As presented in [10], effective circuit cutting must preserve the integrity of quantum information across the divided sections while minimizing the overhead introduced by recombining the results. Addressing these optimization and circuit cutting challenges is essential for the practical deployment of adaptive quantum circuits, necessitating ongoing research and innovative solutions to fully harness their potential.

### B. Related works

Adaptive quantum circuits have been explored to enhance the flexibility and efficiency of quantum algorithms. Studies by GRIMSLEY et al. [9] and MAGNUSSON et al. [8] introduced adaptive ansatz construction techniques that dynamically adjust the circuit based on intermediate measurements, leading to reduced circuit depth and improved performance on NISQ devices. These flexible strategies have demonstrated significant potential across a variety of fields, particularly in quantum chemistry and machine learning, highlighting the potential of adaptive approach for quantum circuits.

The concept of distributed quantum computing, where quantum tasks are divided across multiple QPUs, has gained traction as a strategy to overcome hardware limitations and enhance computational scalability [11]. Several research have focused on circuit partitioning [12][13] and error mitigation techniques [14] essential for effective multi-QPU deployments. Hypergraphic partitioning methods [15][16], which involve breaking down complex circuits into smaller, manageable subcircuits, have been proposed to optimize resource allocation and minimize communication overhead.

*C. Paper organization*

This paper is organized as follows. Section II introduces adaptive quantum circuits, highlighting how they dynamically differ from traditional designs. Section III presents the hypergraphic representation, starting with static circuits in Subsection A, then expanding to adaptive circuits in Subsection B, introducing a new partitioning algorithm in Subsection C, and comparing static and adaptive methods in Subsection D. Section IV covers experiments and results, including benchmarks, methods, an illustrative example, and the benefits of hypergraphs for circuit cutting. Section V concludes by summarizing key contributions and proposing future research directions.

## II. ADAPTIVE QUANTUM CIRCUITS

*A. Static versus adaptive quantum circuits*

Static quantum circuits use a fixed sequence of gates that remains unchanged during execution, ensuring predictability and simplicity. This rigidity suits well-understood tasks, enabling efficient gate placement and minimal circuit depth.

A static circuit $C$ can be represented as:

$$C = U_N U_{N-1} \dots U_2 U_1 \quad (1)$$

where $U_i$ are unitary operations (quantum gates) applied sequentially on the quantum state $|\psi\rangle$.

In contrast, adaptive quantum circuits, also called dynamic circuits [5] or hybrid circuits [6], incorporate adaptive elements that allow for modifications based on intermediate measurements or computational feedback during execution. These circuits can alter their structure in real-time, introducing conditional operations, dynamic gate insertions, or parameter adjustments based on the outcomes of measurements performed on ancillary qubits or in specific parts of the circuit.

A dynamic quantum circuit $C'$ can be expressed as a sequence of conditional operations:

$$C' = (U_N|m_N\rangle)(U_{N-1}|m_{N-1}\rangle)\dots(U_2|m_2\rangle)(U_1|m_1\rangle) \quad (2)$$

where $|m_i\rangle$ represents the measurement outcomes that influence the subsequent gate operations $U_{i+1}$.

Figure 1 illustrates a static circuit (a) and an adaptive circuit (b), also called dynamic circuit in several papers and platforms.

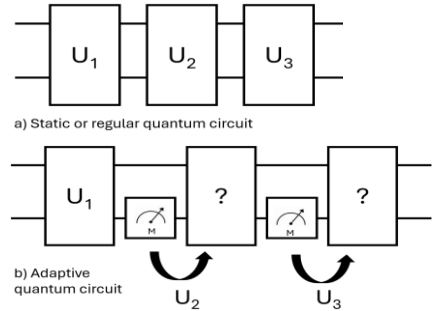

*Figure 1. Static quantum circuit (a) with fixed set of gates and adaptive quantum circuit (b) with flexible set of gates in runtime, based on intermediate measurements $|m_i\rangle$.*

Static quantum circuits are simpler to design and analyze but lack the flexibility to adapt to intermediate computational states. In contrast, adaptive circuits, though more complex, offer enhanced resource efficiency [8], real-time error mitigation [17], and algorithmic flexibility. They dynamically allocate quantum resources, reducing unnecessary gate operations and qubit usage, and use real-time feedback to implement on-the-fly error correction, thereby improving fidelity and convergence. However, these benefits come with challenges. The increased complexity of adaptive circuits demands sophisticated control mechanisms and real-time decision-making [18], and the need for immediate feedback can introduce latency [19]. Furthermore, current quantum hardware limitations, such as gate fidelity and coherence time constraints, add additional hurdles to the practical implementation of dynamic circuits.

IBM's Heron series is an example of an ISA for adaptive quantum circuits, featuring 133 qubits and an adjustable coupler architecture [20]. Similarly, Quantinuum recently introduced its H1 Series, which supports adaptive behavior through the measurement-based quantum computation (MBQC) framework [22]. Additionally, [23] presents a scalable framework that leverages the hybrid capabilities of quantum and classical processors.

Using this adaptive approach, figure 2 illustrates an example of quantum circuit, as defined in [26], with adaptive constructions, such as "WHILE" and "IF/ELSE".

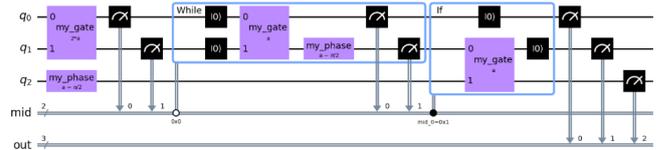

*Figure 2. A adaptive quantum circuit with "while" and "if..else" instruction after intermediate measurement.*

The adaptive circuit from figure 2 applies "my_gate" and "my_phase" operations to qubits $q_0$ and $q_1$, then performs an intermediate measurement on $q_0$. The measurement outcome drives a "WHILE" loop that repeatedly applies "my_gate" and "my_phase" until the loop condition is no longer met. A subsequent measurement on $q_0$ triggers an "IF" statement, which conditionally applies another "my_gate" on $q_0$. The circuit concludes with final measurements on all qubits, capturing the outcomes of these adaptive steps.

*B. Challenges for optimization and circuit partitioning*

Optimizing static quantum circuits already presents challenges [27], but adaptive circuits add further complexity due to their dynamic nature. Managing conditional operations based on real-time measurements requires integrated quantum-classical control and synchronized communication across multiple QPUs. Effective partitioning is crucial to minimize inter-QPU communications that can cause delays and degrade performance. Moreover, the inherent adaptivity of these circuits renders static analysis and traditional optimization methods insufficient, while NISQ devices exacerbate issues such as gate errors and decoherence. Determining an optimal partitioning strategy that considers both logical gate groupings and dynamic adaptivity remains a significant challenge impacting overall performance and reliability.

### III. Hypergraphic representation

In the realm of quantum circuits, a graph (G) is used to represent the system where each vertex corresponds to a qubit and each edge signifies a two-qubit gate, such as CNOT, CZ, CP, or SWAP [12]. Expanding on this, a hypergraph (H) serves as a more versatile structure that can handle more complex scenarios within quantum circuits. Unlike regular graphs, hypergraphs allow edges to connect multiple vertices simultaneously, making them ideal for representing multi-qubit gates like CCNOT or CSWAP, as well as tripartite entanglement correlations, quantum error correction codes, and intricate state configurations.

Quantum circuits can be effectively modeled using primal hypergraphs (H = (V, E)), where (V) represents the set of qubits and (E) denotes the multi-qubit gates. To optimize the execution of these circuits, (H) can be divided into smaller subgraphs ($H_1$, $H_2$,..., $H_k$). This partitioning enables each subcircuit to operate on separate quantum processing units (QPUs), enhancing overall efficiency. The task of partitioning is an optimization problem that aims to minimize specific objectives through heuristic approaches. Commonly used methods for this purpose include the Kernighan-Lin algorithm [24] for graph partitioning and the Fiduccia-Mattheyses algorithm [25] for hypergraphs. In our study, we applied the Kernighan-Lin algorithm alongside a modified version of the Fiduccia-Mattheyses heuristic, taking into account both weights and balance constraints to achieve optimal partitioning.

This structured approach allows for more manageable and efficient execution of quantum circuits, leveraging the strengths of hypergraphic representations and advanced partitioning algorithms to overcome the complexities inherent in quantum computing.

#### A. Hypergraphic representation for static circuits

In static circuits, each gate is modeled as a hyperedge connecting the relevant qubits. To facilitate partitioning and mapping onto multi-QPU systems, we represent a static quantum circuit as a hypergraph using the pseudo-code in Algorithm 1.

*Algorithm 1. Translating static circuits into hypergraph representation.*

```
01  Algorithm StaticCircuitToHypergraph(circuit):
02    Input: circuit - A static quantum circuit with gates and qubits
03    Output: H - Hypergraph representing the circuit
04    Initialize hypergraph H = (V, E)
05    V = set of qubits in the circuit
06    E = empty set
07    for each gate g in circuit:
08      q_list = list of qubits that g operates on
09      Create hyperedge e_g connecting vertices in q_list
10      Assign weight w(e_g) based on gate properties
11      Add e_g to E
12    return H
```

Algorithm 1 converts a static quantum circuit into a hypergraph for analysis and partitioning. It begins by initializing an empty hypergraph with vertices representing the qubits (lines 4-6). Then, for each gate, it creates a hyperedge connecting the qubits it acts on and assigns a weight based on properties like duration or error rates (lines 7-11). Finally, the hypergraph is returned (line 12), capturing the qubit relationships for further processing, such as partitioning for multi-QPU execution.

In algorithm 1 we have:

- Vertices (V): each qubit in the circuit becomes a vertex in the hypergraph.
- Hyperedges (E): each gate is represented as a hyperedge connecting all qubits it operates on.
- Weights (w): hyperedges may be assigned weights reflecting the cost or importance of cutting that gate during partitioning.

Using this approach, benchmark circuits can be generated in hypergraph form. An incidence matrix, whose dimensions match the number of vertices and edges (e.g., a 4×6 matrix for a hypergraph with 4 vertices and 6 edges), captures the relationships between vertices and hyperedges. In the matrix, each row represents a vertex and each column a hyperedge, with entries of 1 indicating a connection and 0 indicating none. This structure clearly visualizes the hypergraph's topology, and a value of -1 can denote specific constraints, such as grouping certain edges or gates from the original circuit.

#### B. Hypergraphic representation for adaptive circuits

To represent adaptive circuits with hypergraphs, we incorporate additional constructs that capture their dynamic aspects. Conditional hyperedges are included or omitted based on measurement outcomes or classical computation results, while temporal layers provide a time-indexed view of the circuit at different execution stages, reflecting its evolving structure. Using this approach, the algorithm to translate the adaptive circuit into hypergraph representation is presented in algorithm 2.

*Algorithm 2. Translating adaptive circuits into hypergraph representation.*

```
01  Algorithm AdaptiveCircuitToHypergraph(adaptive_circuit):
02    Input: adaptive_circuit - An adaptive quantum circuit with gates, qubits, and classical controls
03    Output: H - Hypergraph representing the adaptive circuit

04    Initialize hypergraph H = (V, E)
05    V = set of qubits in adaptive_circuit
06    E = empty set

07    for each layer t in adaptive_circuit:
08      for each gate g in layer t:
09        q_list = list of qubits that g operates on
10        if g is a conditional gate:
11          C_g = condition for gate g (based on meas. outcomes)
12          Create conditional hyperedge e_g with condition C_g
13          Assign weight w(e_g) based on gate properties
14        else:
15          Create hyperedge e_g
16          Assign weight w(e_g) based on gate properties
17        Add e_g to E

18      for each measurement m in layer t:
19        q_m = qubit being measured
20        if m influences future gates:
21          Create measurement hyperedge e_m connecting q_m
22          Assign weight w(e_m) based on its impact on circuit
23          Add e_m to E
```

```
24    // Grouping gates into logical modules
25    for each group G:
26        Create super-hyperedge E_G connecting all qubits in G
27        Assign weight w(E_G) as the sum of w of hyperedges in G
28        Remove individual hyperedges in G from E
29        Add E_G to E

30    return H
```

Algorithm 2 converts an adaptive quantum circuit into a hypergraph. It starts by initializing an empty hypergraph where vertices represent qubits and edges represent gates and measurements (lines 4–6). The circuit is then processed layer by layer; for each gate, if it is conditional, a corresponding conditional hyperedge is created (lines 7–17), while standard gates are represented with weighted hyperedges based on their properties. Measurements that influence subsequent operations are also captured with dedicated hyperedges (lines 18–23). To reflect the circuit's logical structure and simplify the hypergraph, gates are grouped into logical modules, and super-hyperedges are formed (lines 24–29). Finally, the complete hypergraph is returned (line 30) for further use in partitioning and distributing the circuit across multiple QPUs.

This approach defines key properties for modeling adaptive quantum circuits. Vertices (V) represent qubits, while Hyperedges (E) capture circuit operations, categorized as standard (always executed), conditional (dependent on measurement outcomes), or measurement-based (influencing future gates). Super-Hyperedges group related gates into modules to enhance partitioning efficiency. And Weights (w) reflect the cost of cutting hyperedges, factoring in execution probability and adaptivity. This method effectively models essential features of adaptive circuits, including logical execution groups, gate parallelism, conditional operations, and isolated binary gates.

### C. Novel hypergraphic partitioning algorithm for adaptive quantum circuits

Partitioning the hypergraph of an adaptive quantum circuit into balanced subsets suitable for execution on multiple QPUs aims to minimize inter-QPU communication while preserving adaptive behaviors. In our new hypergraph approach, logical constructions are represented as hyperedges, and classical hypergraph heuristics like the Fiduccia-Mattheyses (FM) algorithm [40] remain effective for the partitioning process. The FM algorithm is particularly well-suited for real-time constraints due to its superior performance and flexibility, operating as a linear-time heuristic relative to the number of vertices.

In this context, algorithm 3 presents our new method for hypergraph partitioning adapted specifically for adaptive quantum circuits. Based on the Fiduccia-Mattheyses heuristic, we enhance the algorithm by adding components that capture the logical groups and conditional hyperedges present in the original circuit.

*Algorithm 3. Translating adaptive circuits into hypergraph representation.*

```
01  Algorithm AdaptiveCircuitPartitioning(H, k):
02  Input: H - Hypergraph of the adaptive circuit
03         k - Number of partitions (number of QPUs)
04  Output: P - Partitioning of H into k subsets (P = {V1, V2, ..., Vk})

05  // Initial Partitioning
06      Initialize P using a simple heuristic (e.g., cutting the circuit in the middle for width dimesion)

07  // Apply Hypergraph Partitioning Heuristic (e.g., Fiduccia-Mattheyses Algorithm)
08  repeat until convergence:
09      for each vertex v in V:
10          for each neighboring partition Vi:
11              Calculate gain G(v, Vi) of moving v to Vi
12              G(v, Vi) = ΔCutSize(v, Vi) - λ * ΔBalance(v, Vi)
13              ΔCutSize(v, Vi): Change in cut size if v is moved to Vi
14              ΔBalance(v, Vi): Change in balance metric
15              λ: To control trade-off between cut size and balance
16          Select partition Vi* that maximizes G(v, Vi)
17          if G(v, Vi*) > 0:
18              Move v to partition Vi*
19              Update partition assignments
20              Update gains for affected vertices

21  // Handle Conditional Dependencies
22  for each conditional hyperedge e_c in E:
23      if e_c is cut between partitions:
24          Insert communication protocols to handle
                    conditional executions across QPUs
26          Adjust weights w(e_c) to reflect
                    communication overhead
28  return P
```

Algorithm 3 partitions an adaptive quantum circuit, represented as a hypergraph, across multiple QPUs to balance workload, reduce inter-QPU communication, and preserve circuit adaptivity. The process starts with an initial vertex partitioning using a simple heuristic (line 6). It then employs an iterative approach based on the Fiduccia-Mattheyses heuristic (lines 8–20), where vertices are moved to neighboring partitions if doing so yields a positive gain that improves both cut size and partition balance (lines 12–15). After this, the algorithm addresses conditional dependencies by detecting cut conditional hyperedges (lines 22–27) and inserting communication protocols while adjusting weights for added overhead. Finally, the partitioning result is returned (line 28).

The overall process aims to optimize the distribution of quantum circuit components across QPUs while addressing both computation and communication efficiency.

### D. Example illustration

Consider the following static circuit with 3 qubits and 2 binary gates, as illustrated in figure 3.

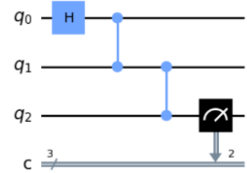

*Figure 3. Static circuit for example illustration.*

From the static circuit represented in figure 3. we have the following components:

- *Qubits: $q_0$, $q_1$, $q_2$*
- *Gates: $G_1$ (on $q_0$, $q_1$), $G_2$ (on $q_1$, $q_2$)*

The hypergraph Representation can be described as:

- *Vertices: V = {$q_0$, $q_1$, $q_2$}*
- *Hyperedges: E = {$e_1$ ($G_1$ connecting $q_0$, $q_1$), $e_2$ ($G_2$ connecting $q_1$, $q_2$)}*

Now, let´s consider the following adaptive quantum circuit, with a conditional gates "if" with impact for qubits $q_1$ and $q_2$, as illustrated in figure 4.

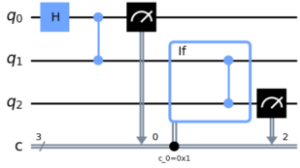

*Figure 4. Adaptive quantum circuit for example illustration.*

From this adaptive quantum circuit in figure 4, we have the following components:

- *Qubits: $q_0$, $q_1$, $q_2$*
- *Gates:*
  - *$G_1$ (on $q_0, q_1$)*
  - *$G_c$ (conditional gate on $q_1, q_2$) executed if $M(q_0) == 1$*
- *Measurement: $M(q_0)$*

The hypergraph representation for this adaptive quantum circuit can be described as:

- *Vertices: $V = \{q_0, q_1, q_2\}$*
- *Hyperedges:*
  - *$e_1$: Hyperedge for $G_1$ connecting $q_0, q_1$*
  - *$e\_m$: Measurement hyperedge for $M(q_0)$ connecting $q_0$*
  - *$e\_c$: Conditional hyperedge for $G_c$ connecting $q_1, q_2$ with condition $C\_gc$ ($M(q_0)=1$)*
- *Weights:*
  - *$w(e1)$: Based on $G_1$ properties*
  - *$w(e\_m)$: Based on measurement impact*
  - *$w(e\_c)$: Weighted by the probability that $M(q_0)=1$*

Comparing these two approaches, two different impacts for the partitioning process are noted:

- First, in the classical approach for static circuits, we may cut e_c without considering its conditional nature, potentially leading to high communication overhead if Gc is frequently executed.
- Second, in the novel approach for adaptive quantum circuit, we recognize the conditional dependency between $M(Q_0)$ and $G_c$. Because of that, we keep $Q_1$ and $Q_2$ in the same partition if $M(Q_0)$ is likely to be 1, reducing inter-QPU communication for $G_c$.

Now, let´s combine more gates in a bigger circuit as example. Figure 5 illustrate an adaptive quantum circuit for 3 qubits and 2 classical bits, two support intermediate measurements and outputs.

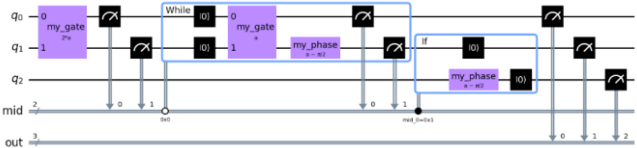

*Figure 5. Static circuit for example illustration.*

From figurw 5 we can represent the various components of this circuit with attributes like Qubits, Gates, Measurements, Vertices, Hyperedges, and Weights. This approach gives a structural perspective of the circuit in terms of hypergraph representations, emphasizing both the classical and quantum components of the adaptive quantum circuit.

Table 1 illustrates the main components from this adaptive circuit, considering the hypergraph representation approach proposed in this paper:

*Table 1. Main components from the adaptive circuit in a hypergraph representation approach.*

| Components | Description |
|---|---|
| Qubits (Vertices) | $V = \{q_0, q_1, q_2\}$ |
| Classical Bits (Vertices) | classical vertices include mid[0], mid[1], out[0], out[1], out[2] |
| Gates (Hyperedges) | $e_1$: my_gate(a * 2) on $q_0, q_1$ <br> $e_2$: my_phase(a) on $q_1$ |
| Measurements (Hyperedges) | $em_0$: Measure $q_0$ to mid[0] <br> $em_1$: Measure $q_1$ to mid[1] |
| Control Flow (Hyperedges) | $e_{while}$: while (mid == "00") block <br> $e_{if}$: if (mid[0]) block |
| Reset Operations: | $er_0$: Reset $q_0$ <br> $er_1$: Reset $q_1$ <br> $er_{inner}$: Reset *inner_alias* |
| Weights (W) | Each hyperedge is assigned a weight based on gate properties, measurement impacts, or control flow conditions |

The hypergraph representation helps visualize and understand the relationships and dependencies in a quantum circuit by treating qubits as vertices and operations (gates, measurements, control flow) as hyperedges that connect these vertices. This allows for a structured analysis of both static and adaptive quantum circuits, providing insights into:

- Gate Connectivity: How gates connect different qubits, represented through hyperedges connecting the corresponding vertices.
- Measurement and Control Flow: Adaptive quantum circuits involve conditional gates and measurements that introduce classical feedback. These can be represented through special conditional hyperedges and measurement hyperedges that show the influence of classical bits on quantum operations.
- Weight Assignments: Assigning weights to hyperedges allows for modeling quantum properties, such as the likelihood of conditions being true, the influence of measurements, and gate complexities. This can help in optimizing quantum circuits, analyzing error propagation, and understanding circuit depth and runtime.

*E. Comparison between static and adaptive approaches*

The hypergraph representation differs significantly between static and adaptive quantum circuits. In the static approach, only fixed gates and qubits are represented, with each gate treated as independent hyperedge. In contrast, the adaptive approach incorporates conditional gates and measurements such as hyperedges, grouping logically connected gates into super-hyperedges to enhance partitioning efficiency. Weight assignments also vary static circuits use fixed gate properties, while adaptive circuits assign dynamic weights based on execution probabilities and adaptivity. Partitioning strategies further distinguish the two, as static circuits rely on standard partitioning methods, whereas adaptive circuits adjust gain calculations and movement strategies to minimize communication overhead. By accounting for circuit dynamics, the adaptive approach improves efficiency, reducing inter-QPU communication and maintaining workload balance.

## IV. EXPERIMENTS AND RESULTS

### A. Methdology

To assess the impact of hypergraphic representation on circuit cutting in adaptive quantum circuits, we employ a systematic methodology. First, we model both static and adaptive circuits as hypergraphs, where vertices represent qubits and hyperedges capture multi-qubit interactions, thus reflecting the complex dependencies of quantum operations. We then convert these hypergraphs into incidence matrices, which serve as input for various partitioning heuristics. Partitioning algorithms, including Kernighan-Lin and a modified Fiduccia-Mattheyses, are applied to divide the hypergraphs into smaller subgraphs. For static circuits, the focus is on minimizing inter-subgraph interactions, while for adaptive circuits, partitioning also accommodates dynamic adjustments from intermediate measurements. Finally, the subgraphs are distributed across multiple QPUs, with performance evaluated in terms of qubit reuse, error rates, and overall computational fidelity.

### B. Benchmark circuits

For adaptive circuits, we selected the following list of algorithms:

- Variational Quantum Eigensolver (VQE) uses parameterized circuits with classical feedback to find a Hamiltonian's ground state energy, making it well-suited for HyPAQ's adaptive partitioning that efficiently handles gate groupings and dependencies [28].

- Quantum Phase Estimation (QPE) estimates a unitary operator's eigenvalues via a quantum register and the Quantum Fourier Transform, with qubit requirements scaling with precision [29].

- Iterative Quantum Phase Estimation (IQPE) achieves similar estimates using only two qubits by repeatedly applying controlled operations and updating phase estimates, a process that fits well with hypergraph-based dependency management [30].

- Random circuits, which combine randomly selected gates with classical instructions like "FOR", "WHILE", "IF", and "ELSE", further illustrate adaptive behavior [26].

- Additionally, the Generalized Repeat-Until-Success (RUS) circuit for Rz(θ), as detailed in Nielsen and Chuang [31], implements an iterative Z-rotation protocol (with $\cos(\theta-\pi) = 3/5$) and scales from 4 to 48 qubits, exemplifying the adaptive processes leveraged by HyPAQ.

For all the experiments, we used Python 3 with Qiskit 1.2 [32] platform for the quantum circuits implementations. We used OPENQASM 3 [33] to support classical instructions for the adaptive quantum circuit approach. For simulations and partitioning experiments we used a local machine with 32 GB RAM of memory, 12th Gen Intel® Core ™ i7 1270p, 2200 Mhz, 12 Core(s), and 16 Logical Processors.

### C. Experiments and results

Figure 6 presents the complexity of each circuit, captured by groups of hyperedges, instead of classical hyperedges based only on binary or multi qubit gates.

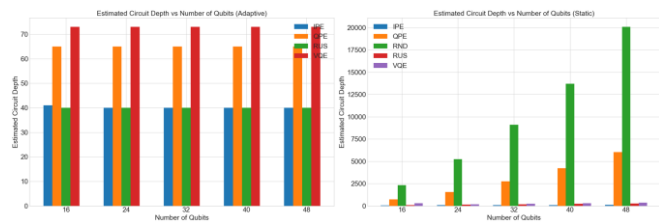

*Figure 6. Estimated circuit depth vs. Number of qubits for ataptive and static circuits, captured from primal hypergraph (static) and extended hypergraph (adaptive)*

Our approach groups qubits and operations within adaptive circuits, prioritizing local subgraphs over cross-partition splits. This results in hypergraph with additional elements representing measurements, classical instructions, and gate groupings. However, partitioning may lead to imbalance due to the inclusion of these grouped classical operations.

Figure 7 presents the impact of hypergraph representation for each static and adaptive circuit, based on the number of gates versus number of qubits. .

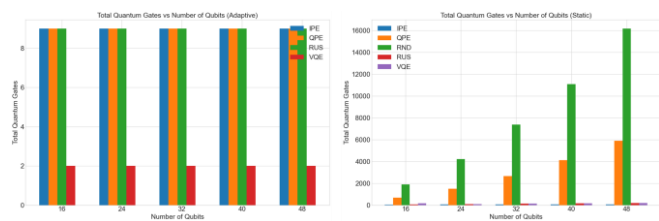

*Figure 7. Total quantum gates vs. Number of qubits for adaptive and static circuits, using hypergraph representation.*

The hypergraphic representation of adaptive quantum circuits offers key advantages, including sparsity and reduced complexity, which streamline circuit cutting by minimizing interdependencies. Unlike traditional graphs, hypergraphs maintain predictable complexity across varying circuit sizes, making them ideal for adaptive circuits with dynamic control structures like "IF" and "WHILE." This efficiency enhances decomposition and execution in distributed quantum environments.

## V. DISCUSSIONS AND CONCLUSIONS

In this research, we introduced a novel algorithm that maps adaptive quantum circuits into hypergraph representations, diverging from traditional approaches that translate static circuits for spatial circuit cutting. By incorporating adaptive elements into the hypergraph representation, we addressed the challenges posed by dynamic circuit structures in adaptive quantum algorithms. The findings from this approach indicate that considering the adaptivity of quantum circuits in their hypergraph representation and partitioning is crucial for optimizing the execution of quantum algorithms on distributed architectures. This advancement is particularly significant for complex applications, where circuit depth and qubit requirements exceed the capacity of single QPU systems.

### A. Contributions and future work

This research introduces a novel hypergraph representation for adaptive quantum circuits, incorporating conditional hyperedges and measurement dependencies to model quantum operations and control flow. We extend circuit cutting techniques to dynamic circuit structures, enabling efficient partitioning and distribution across multiple QPUs while preserving adaptivity. Additionally, we optimize hypergraph heuristics by adapting partitioning algorithms like

Fiduccia-Mattheyses, grouping gates into hyperedges, and adjusting weights based on adaptivity and execution probabilities to improve efficiency and convergence.

Future work will focus on scaling this framework for larger circuits with more qubits and deeper adaptive layers, ensuring practical viability. Integration with quantum hardware and high-fidelity simulators will address real-world constraints like qubit connectivity, gate fidelities, and decoherence. Further optimization of inter-QPU communication - through compression, asynchronous protocols, and error mitigation - will minimize execution overhead in distributed environments. Finally, we aim to extend these methods to adaptive algorithms such as QAOA and quantum simulations, broadening their impact on distributed quantum computing.